\PassOptionsToPackage{hyphens}{url}
\documentclass[final,5p,times,twocolumn]{elsarticle}

\usepackage{graphicx}      
\usepackage{natbib}        
\usepackage{amsmath}  
\usepackage{amssymb}
\usepackage{mathtools}
\usepackage{booktabs}
\usepackage{xcolor}
\usepackage{subcaption}
\usepackage{xurl}
\usepackage[hyphens]{url}

\newtheorem{remark}{Remark}

\begin{document}
\begin{frontmatter}

\title{Recurrent Convolutional Neural Networks for LiDAR-Based  {Attitude} Initialization of Rotating Spacecraft} 

\author[First]{Luca Bechis}\ead{s315994@studenti.polito.it} \author[First]{Jean-Luc Sarvadon}\ead{jeanluc.sarvadon@polito.it} \author[First]{Petre Ricioppo}\ead{petre.ricioppo@polito.it} \author[First]{Mauro Mancini\corref{cor1}}\ead{mauro.mancini@polito.it} 
\affiliation[First]{organization={Department  of Mechanical and Aerospace Engineering, Politecnico di Torino}, 
addressline={Corso Duca degli Abruzzi 24},  
city={Torino}, 
postcode={10129}, 
country={Italy}} 
\cortext[cor1]{Corresponding author} \fntext[]{Received 2 November 2025, Revised 20 June 2026, Accepted 20 June 2026 \newline Digital Object Identifier (DOI): \url{10.1016/j.conengprac.2026.107134}}

\begin{abstract}  
Accurate  {attitude} estimation is essential for autonomous in-orbit servicing and proximity operations. This work proposes a Recurrent Convolutional Neural Network (RCNN) used in  {coarse attitude initialization} of known, possibly tumbling spacecraft using LiDAR-derived depth images. By processing temporal sequences of 2D point-cloud projections, the RCNN effectively handles symmetries, occlusions, and degraded sensing. Simulations across various spacecraft geometries, angular velocities, and ranges show that the RCNN yields lower initialization errors and higher convergence rates than conventional CNN  {baseline within the adopted experimental framework, with performance varying across angular velocity conditions.}

\end{abstract}


\end{frontmatter}

\textcopyright 2026. This manuscript version is made available under the CC BY 4.0 license \url{https://creativecommons.org/licenses/by/4.0/}

\section{Introduction}
In recent years, space agencies, private companies, and the research community have shown a growing interest in In-Orbit Servicing (IOS) missions, including satellite inspection, capture, refueling, repair, and debris removal. These operations play a key role in reducing mission costs and extending spacecraft lifetimes, paving the way toward a more sustainable and accessible use of space \citep{OPROMOLLA2024469}.

Despite its promising potential, IOS missions face several engineering challenges, mainly related to the design of modular and reconfigurable spacecraft, the implementation of high-precision propulsion systems, and the realization of advanced Guidance, Navigation, and Control (GNC) architectures capable of supporting fully autonomous operations in complex orbital environments. Among them, the GNC subsystem is particularly critical, as it is responsible for precise relative motion estimation, trajectory planning, and control during close-proximity operations. Its performance directly affects the ability of the servicing spacecraft to safely approach, capture, and manipulate client satellites, especially when these are non-cooperative or unprepared for docking \citep{MOGHADDAM202170}.  
In particular, the relative navigation function in IOS missions is responsible for relative state estimation between chaser and target, i.e., for determining the relative position, velocity, attitude, and angular velocity \citep{Inorbitservicing}.
The main sensors involved in pose estimation are LiDAR and cameras \citep{Fehse_2003, OPROMOLLA4}, 
which offer complementary advantages: LiDAR provides accurate geometric information and operates under any lighting condition, whereas cameras deliver higher-resolution imagery but rely on favorable illumination for effective 3D reconstruction.
In LiDAR-based pose estimation of known targets, the pose is determined by aligning the scanned point cloud with a reference point cloud representing the target, hereafter referred to as the model, through rotation and translation.
The process typically involves two phases: a \textit{coarse phase}, which provides a rough but robust initial pose estimate, followed by a \textit{fine phase}, where the estimate is refined through a point cloud registration method. Among the pose refinement techniques proposed in the literature \citep{OPROMOLLA4}, the most widely employed is the Iterative Closest Point (ICP) algorithm \citep{Besl, yang}. The accuracy of ICP depends on the number of points and measurement noise, and its convergence critically relies on the availability of a sufficiently accurate initial guess.
In order to provide a reliable approximation of the target pose before the refinement stage, various methods have been proposed in the literature.  A comprehensive review of LiDAR-based coarse pose estimation methods can be found in \citep{OPROMOLLA4}, while in the following, some of the most representative approaches are briefly described.

\citep{opromolla2,opromolla3} employ the template matching method, which uses a dataset of templates to optimize a correlation function. This technique allows initial estimates with errors of around 20° for non-symmetrical targets but remains computationally and memory-intensive. 
A different approach is proposed by \citep{yin}, where the initialization estimate is obtained by aligning two congruent tetrahedra constructed on the model and on the scanned point cloud. On that particular shape, the algorithm achieved errors below 9°. 
Another approach is the Point Pair Feature (PPF) method \citep{drost}, which estimates the pose by encoding geometric relationships between pairs of points. These model features are precomputed and stored, while the same features extracted from the scene are used to find correspondences. Then, a voting process selects the transformation that best aligns the two point clouds, yielding robust and accurate results with a computational load that depends on the point cloud density. 
Another feature-based approach is the Oriented, Unique, and Repeatable Clustered Viewpoint Feature Histograms (OUR-CVFH). The pipeline of this algorithm consists of identifying groups of points with common surface features. These features are then compared with histograms obtained from different viewpoints to derive the desired initialization. Exploiting this method, \citep{sell} achieved convergence with errors below 10° on all axes, while \citep{woods} introduced an additional filtering stage that further reduced attitude errors and improved pose tracking performance. 
Although these initial pose estimation approaches offer significant potential, they may fail in symmetrical configurations and can be computationally intensive. The growing deployment of machine learning has enabled alternative methods for coarse pose estimation phase. For instance, \citep{d'amico2} and \citep{d'amico1} achieved encouraging results in satellite pose estimation using Convolutional Neural Networks (CNNs) trained on synthetic images and tested on real camera data. Conversely, \citep{zhang} employed transformers to provide coarse estimates using LiDAR point clouds. Similarly, recent works have explored deep learning approaches for pose estimation directly from LiDAR point clouds. In particular, \citep{Hashimoto} and \citep{Piccinin} proposed strategies targeting short-range scenarios, while \citep{Enrique} developed a neural network specifically designed for operation at a fixed distance of 100 m.
         
As a preliminary step in pose estimation, pose initialization produces two distinct outputs: translation initialization and attitude initialization. These quantities serve as inputs to the subsequent fine estimation phase. Translation initialization can be addressed using established techniques such as centroiding \citep{OPROMOLLA4} and CNN-based approaches \citep{Leo1}. In contrast, attitude initialization remains a more challenging problem, particularly in dynamic target scenarios involving object symmetries and rotational motion. Motivated by prior work on the application of neural networks to pose estimation, this study aims to overcome the challenge of attitude initialization by supplying an initial guess for ICP.
The present study is inspired by the findings of \citep{Leo1,leo2}, who demonstrated that a convolutional neural network trained on a synthetic LiDAR dataset can estimate pose from the projection of a 3D point cloud. 
However, their approach was limited to a particular target and their network is not  able to consider a sequence of  {measurements}.

Indeed, CNNs operate on single frames and do not explicitly account for temporal correlations between successive observations. This limitation may degrade estimation accuracy in scenarios involving spinning, symmetric targets, partial occlusions, or ambiguous visual configurations. To address these issues, 
\citep{Rondao} proposed a Recurrent Convolutional Neural Network (RCNN) with monocular camera images, leveraging its temporal modeling capability to incorporate information from previous inputs for pose estimation.
In this context, the present study extends the concept by implementing a RCNN architecture trained on sequences of LiDAR-derived point clouds, explicitly designed to handle symmetric spinning targets observed at varying ranges. For completeness, a standard CNN with comparable capacity is also developed to enable a direct performance comparison. The input data are processed to generate consistent 2D projections across distances from 20 to 100 m, with combined angular velocity norms up to 20°/s. Extensive evaluations are performed on spacecraft with different geometries and under varying image quality conditions, including occlusions from appendages. Results show that the proposed RCNN provides improved accuracy compared to the CNN baseline and demonstrates competitive performance with respect to state-of-the-art approaches when considering operational range, target symmetries, and occlusion conditions.

This paper is structured as follows: Section \ref{Dataset} describes the procedure implemented to generate the datasets employed for training and testing the neural networks. Section \ref{network} illustrates the architecture of the two types of networks (CNN and RCNN). Then, Section 4 presents the results and Section 5  {draws} some concluding remarks while suggesting possible future developments of the work.

\section{Dataset development for Neural Network training and testing}
\label{Dataset}
This section describes the complete dataset development pipeline for neural network training and testing. It begins with the generation of relative distances and orientations between the chaser and the target, followed by the simulation of LiDAR measurements from the chaser’s perspective. Finally, the point clouds are processed into 2D depth images through projection and voxelization, producing inputs suitable for both CNN and RCNN architectures.

The dataset used in this work is publicly available at
\url{https://zenodo.org/records/18477119}.

\subsection{Definition of orientations and distances}
In order to train both CNN and RCNN effectively, it is necessary to generate a dataset of input images paired with labels describing the true orientation of the Target Reference Frame (TRF) relative to the LiDAR Reference Frame (LRF), in terms of quaternions. The TRF is a body-fixed reference frame attached to the target geometry, while the LRF is centered on the LiDAR, with the boresight axis aligned with the y-axis and the $x-z$ plane perpendicular to it, as in Figs. \ref{fig:3D_scenario}-\ref{fig:lidar_point_of_view}.
Thus, each sample consists of an image and its corresponding label, which are used to evaluate the network’s predictions through an appropriate loss function.
Since the CNN and RCNN architectures operate on different types of input data (individual images and temporal sequences of 24\;s, respectively), the dataset generation process was designed accordingly, as explained below. 
First, target-LiDAR distances expressed in the LRF are generated using a uniform distribution for both datasets, producing random values in the range of 20 to 100 meters on the boresight axis of the instrument. For temporal sequence data, however, the target–chaser distance was kept constant across all frames within a sequence, considering a station-keeping phase.
 Instead, orientations were generated by independently sampling each quaternion component from a zero-mean, unit-variance normal distribution, $\mathcal{N}(0,1)$. 
 {The resulting vectors were then normalized to unit length, yielding quaternions uniformly distributed over $\mathrm{SO}(3)$, due to the isotropic symmetry of the multivariate Gaussian distribution \citep{SHOEMAKE}.}

 {For both CNN and RCNN training datasets, the initial spacecraft orientation is generated using this same sampling procedure, ensuring identical viewpoint distributions. For RCNN training, temporal sequences are then obtained by propagating the initial quaternion according to kinematic evolution rules, as described below.}

\begin{enumerate}
    \item the initial quaternion of each sequence is generated as in the image case;
    \item a constant angular velocity, both in magnitude and direction, is assigned to simulate rotational motion;
    \item the temporal evolution of the target’s orientation is then computed by propagating the quaternion over time according to the assigned angular velocity:
    \begin{equation}
    \mathbf{q}(t) = \left[\cos\left(\frac{\|\boldsymbol{\omega}_t\| \ t}{2}\right),\frac{\boldsymbol{\omega}_t}{\|\boldsymbol{\omega}_t\|}\sin\left(\frac{\|\boldsymbol{\omega}_t\| \ t}{2}\right)\right]\otimes \mathbf{q}_0
    \end{equation}
    where $\boldsymbol{\omega}_t$ is the spinning angular velocity with a norm ranging from 0 to 20 deg/s; $t$ denotes the time instant at which the quaternion is computed, and $\mathbf{q}_0$ is the initial quaternion.
    
\end{enumerate}

The angular velocity components were drawn from a uniform distribution, providing precise control of their amplitude bounds.

\begin{remark}
    The assumed angular velocity $\boldsymbol{\omega}_t$ defines a purely kinematic attitude evolution. While this motion does not comply with Euler’s rotational dynamics, it is sufficient for training, as the RCNN relies solely on visual information for attitude initialization. A physically consistent tumbling case is considered during testing in Section~4.
\end{remark}

This procedure for generating orientations and distances defines the complete relative pose between the chaser and the target at each time instant, as illustrated in Fig.~\ref{fig:3D_scenario}.

\subsection{LiDAR simulator}
\begin{figure}[h]
    \centering
    \includegraphics[width=1\linewidth]{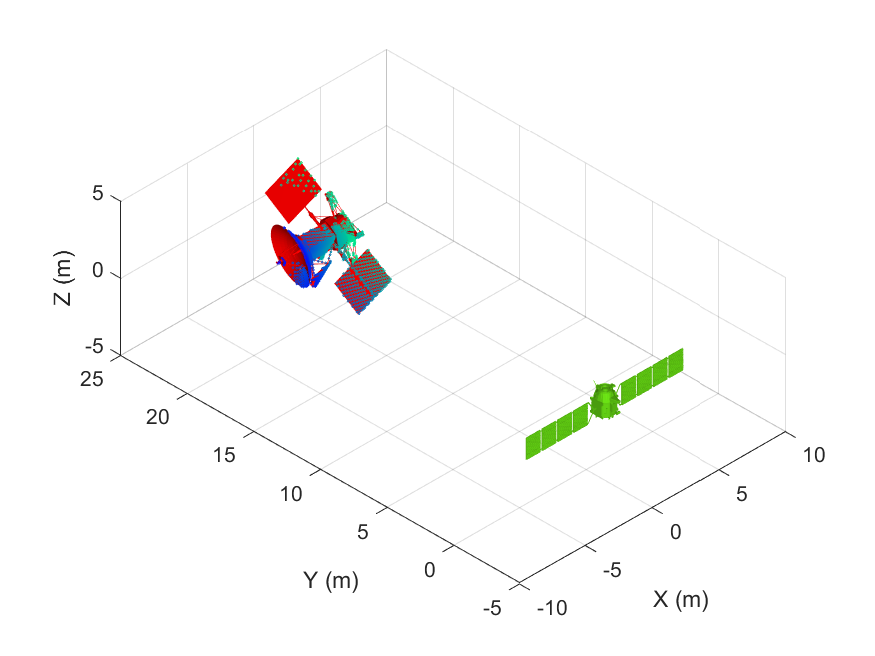}
    \caption{Simulation scenario example. The chaser is depicted in green, while the target is shown in red and illuminated by the LiDAR beam.}
    \label{fig:3D_scenario}
\end{figure}

\begin{figure}[h]
    \centering
    \includegraphics[width=0.7\linewidth]{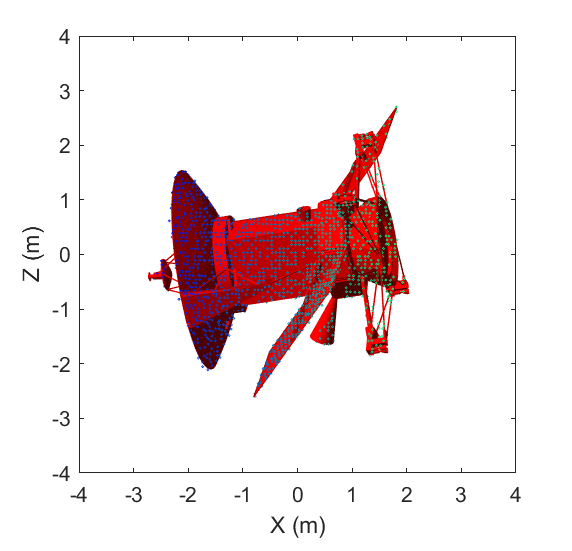}
    \caption{3D view from the LiDAR perspective}
    \label{fig:lidar_point_of_view}
\end{figure}

In this study, the LRF is used to simulate the LiDAR measurements from the chaser’s perspective as shown in the simulation scenario in Fig.~\ref{fig:3D_scenario}. The obtained point clouds are then projected in the plane perpendicular to the boresight axis of the LiDAR (Fig.~\ref{fig:lidar_point_of_view}).
The point clouds were generated using the 3D LiDAR simulator included in Matlab's UAV Toolbox and the CAD models of the target spacecraft, while the models were selected from the resources available on NASA's website \citep{nasa3dresources}.
The main parameters of the considered flash LiDAR are defined in Table \ref{tab:LiDAR_parameters}.
In particular, the sensor noise is defined based on the specifications of the GoldenEye 3D Flash LIDAR\textsuperscript{\texttrademark} Space Camera flown onboard the CST-100 Starliner \citep{Beth}. The corresponding range uncertainty is modeled as zero-mean Gaussian white noise, with a standard deviation $(\sigma$ in Table  \ref{tab:LiDAR_parameters}) equal to the sensor’s specified accuracy, and is added to the nominal range measurement, consistent with common practice in the literature. \citep{opromolla3, Leo1,Gonzales}.

The swath is assumed to be square and large enough to fully encompass the target, while the sensor's Field of View (FOV) is derived as follows:
\begin{equation}\label{eq:FOV}
    \mathrm{FOV} = 2 \cdot \arctan\left(\frac{\mathrm{swath}}{2 \cdot \mathrm{range}}\right).
\end{equation}
In addition, the spacecraft CAD models were selected to ensure that Eq. (\ref{eq:FOV}) gives FOV values between 1° and 40° at the maximum and minimum ranges considered, consistent with commercially available LiDAR sensors.

\begin{table}[h]
    \centering
    \begin{tabular}{cc}
    \toprule
         \textbf{Parameter} & \textbf{Value}\\
         \midrule
         AzimuthLimits & [-FOV/2, FOV/2]\\
         ElevationLimits & [-FOV/2, FOV/2]\\
         AzimuthResolution & FOV/128, FOV/100\\
         ElevationResolution & FOV/128, FOV/100\\
         MaxRange & $500\,\mathrm{m}$\\
         RangeAccuracy ($\sigma$) & $0.1\,\mathrm{m} + 0.01 \cdot \mathrm{range} \,\, $\\
         HasNoise & true\\
         Sample Time & 2.7\,s\\
    \bottomrule
    \end{tabular}
    \vspace{7pt}
    \caption{Selected LiDAR parameters}
    \label{tab:LiDAR_parameters}
\end{table}

For further details on the parameters listed in Table \ref{tab:LiDAR_parameters}, the reader can refer to the dedicated page on the Mathworks website \citep{MathWorks}.

\subsection{Point cloud projection and voxelization}
In order to generate inputs suitable for the neural network, the LiDAR point cloud (Fig. \ref{fig:lidar_point_of_view}) is converted into a depth image through a voxelization step followed by a normalization step. This preprocessing strategy is adopted to limit computational burden and to promote numerical stability, while improving the consistency of point cloud-derived images across varying distances.

\begin{figure}[h]
    \centering
    \begin{subfigure}[b]{0.4\textwidth}
        \centering
        \includegraphics[width=\textwidth]{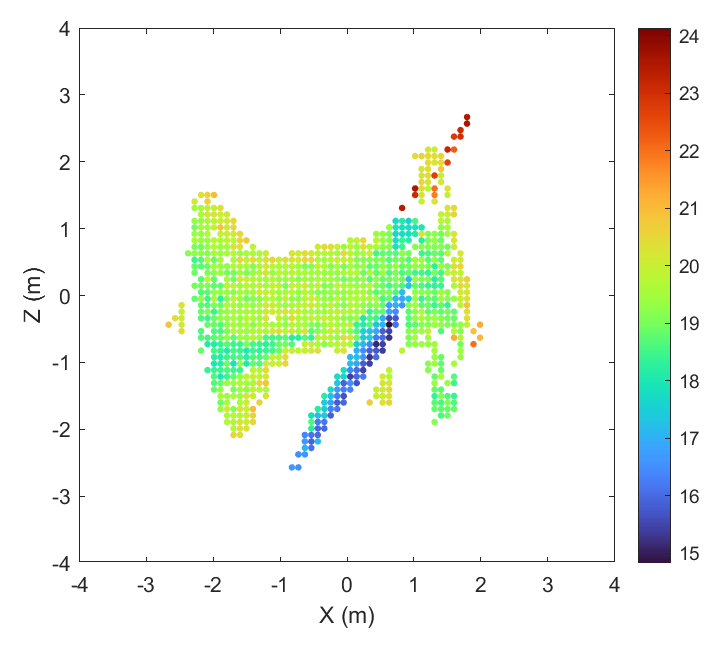}
        \caption{Projected and voxelized point cloud}
        \label{fig:projection_and_voxelization}
    \end{subfigure}
    \quad
    \begin{subfigure}[b]{0.4\textwidth}
        \centering
        \includegraphics[width=\textwidth]{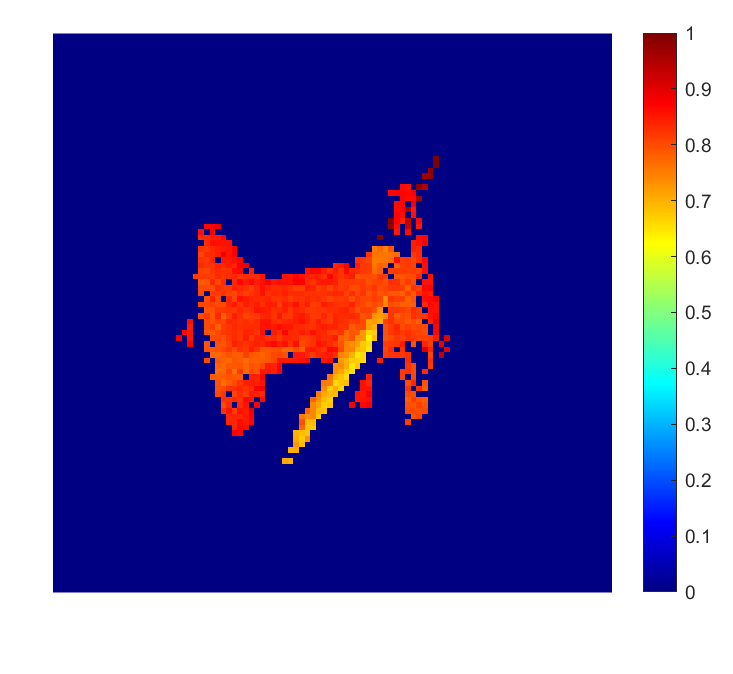}
        \caption{2D depth image}
        \label{fig:2D_depth_image}
    \end{subfigure}
    \caption{Example of transformation from projected point cloud to depth image}
    \label{fig:point_cloud_projection}
\end{figure}

Voxelization consists in partitioning the spatial domain of the point cloud  into a three-dimensional grid of voxels (Fig.~\ref{fig:projection_and_voxelization}). Each voxel is assigned a distance value, whereas cells with no available depth information are assigned a null value. The depth values are subsequently normalized with respect to the maximum value in each image and converted to single-precision format.

The generated dataset comprises depth images such as the example shown in Fig.~\ref{fig:2D_depth_image}, where the color coding is used solely for visualization purpose. The image size was set to 100$\times$100 pixels based on an optimization analysis conducted during the neural network design phase. As described in Section 2.1,  single-image and temporal-sequence datasets are produced separately, and each is divided into independent training/validation and testing sets. For network development, 90\% of the samples are used for training and 10\% for validation. Then, testing is performed on separate data to ensure that all test samples remain unseen during network development.

\section{Network Architecture and Training}\label{network}

This section presents the neural network architectures and the training strategy adopted for spacecraft attitude estimation from depth images. A CNN is initially presented for single-frame attitude regression. The architecture is then extended to a RCNN to exploit temporal information and improve estimation accuracy in the presence of ambiguous visual configurations due to spacecraft symmetries, degraded sensor resolution and partial occlusions. The subsequent section details the adopted training strategy and the configuration of the dataset.

\subsection{Convolutional Neural Network architecture}
\label{sec:CNN}

\begin{figure*}[t]
    \centering
    \includegraphics[width=1\textwidth]{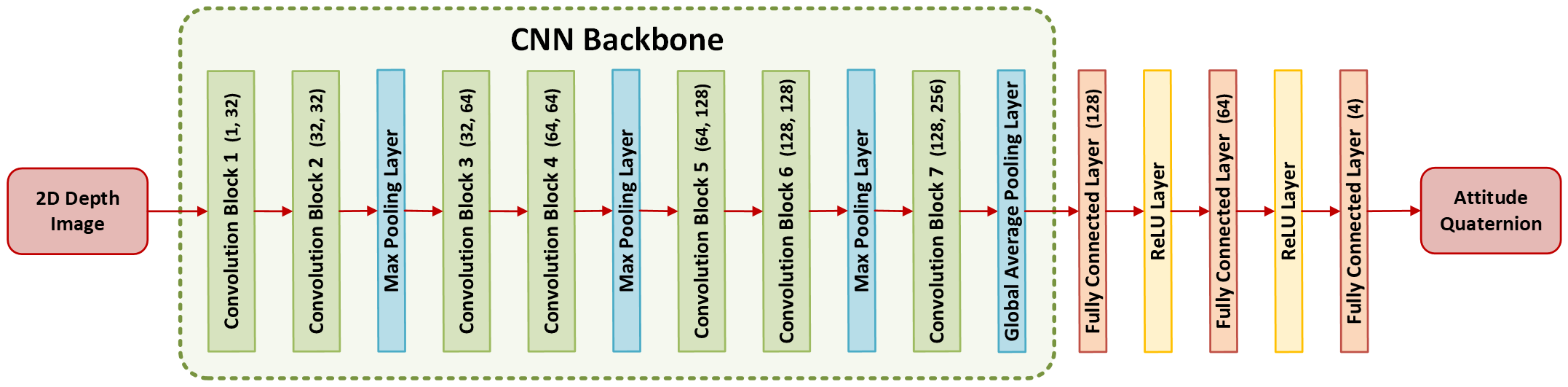}
    \vspace{2mm}
    \refstepcounter{figure}
    \label{fig:CNN_architecture}
    \centerline{%
        \parbox{0.7\textwidth}{%
            \centering
            Fig.~\thefigure. Architecture of the proposed Convolutional Neural Network
        }
    }
\end{figure*}

The architecture of the proposed CNN is provided in Fig. \ref{fig:CNN_architecture}. The network is implemented in Matlab and is designed to achieve an effective trade-off between estimation accuracy and computational complexity. Compared to standard architectures such as ResNet-18, ShuffleNet, and SqueezeNet, the network proposed in this work adopts a relatively shallow structure, resulting in reduced inference times, as discussed in Section~\ref{sec:results}.

The adoption of a compact architecture is driven by the characteristics of the input and the target task. As discussed in~\citep{KECHAGIASSTAMATIS}, depth images primarily encode geometric information at low to intermediate levels, which reduces the need for deep feature hierarchies typically required in RGB-based perception pipelines. Moreover, the limited spatial resolution of the input images constrains the amount of exploitable spatial detail, making a small number of convolutional layers sufficient for effective feature extraction.

\begin{figure*}[t]
    \centering
    \includegraphics[width=0.7\textwidth]{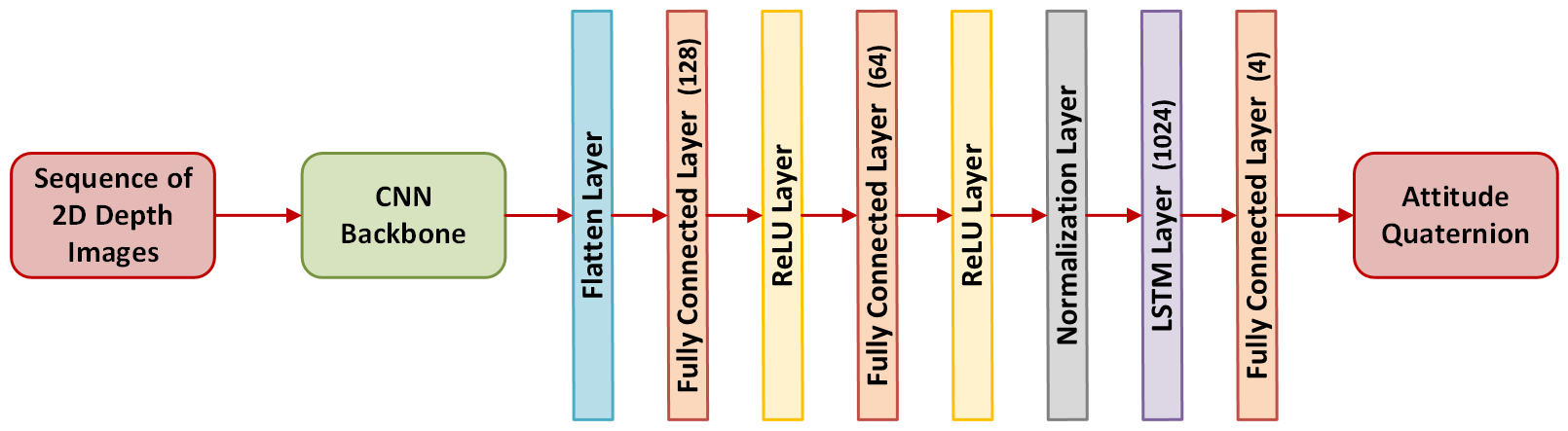}
    \vspace{2mm}
    \refstepcounter{figure}
    \label{fig:RCNN_architecture}
    \centerline{%
        \parbox{0.7\textwidth}{%
            \centering
            Fig.~\thefigure. Architecture of the RCNN.
            The CNN backbone is the same as in Fig.~\ref{fig:CNN_architecture}
        }
    }
\end{figure*}

As shown in Fig.~\ref{fig:CNN_architecture}, feature extraction relies on a sequence of convolutional blocks, while spatial downsampling is explicitly handled through pooling operations. A global average pooling layer is adopted to obtain a compact feature representation, improving robustness to small spatial variations and reducing the number of trainable parameters. The extracted features are processed by fully connected layers with decreasing dimensionality, leading to an output layer that estimates the spacecraft attitude in quaternion form.

The network is trained using a loss function defined in terms of the angular error $\theta$ between the predicted quaternion $\mathbf{q}_{\text{pred}}$ and the ground-truth quaternion $\mathbf{q}_{\text{true}}$, computed as
\begin{equation}
	\theta = 2 \arccos \left( \left| \mathbf{q}_{\text{true}} \cdot \mathbf{q}_{\text{pred}} \right| \right) \cdot \frac{180}{\pi}.
	\label{eq:ang_err}
\end{equation}
The absolute value ensures invariance to the sign ambiguity inherent to quaternion representations, while the training loss is defined as the mean angular error over the batch. Quaternion normalization is enforced both during training and inference to guarantee physically consistent rotation estimates.

\subsection{Recurrent CNN}
\label{sec:RCNN}

The architecture of the proposed RCNN is provided in Fig. \ref{fig:RCNN_architecture}, which shows the integration of Long Short-Term Memory (LSTM) units to model the temporal evolution of the target attitude by exploiting the 
temporal correlations between successive observations.
As shown in Fig.~\ref{fig:RCNN_architecture}, the RCNN extends the CNN by processing short sequences of depth images. Each frame is encoded by the shared CNN backbone described in Section~\ref{sec:CNN}, and the resulting features are fed to a recurrent module based on LSTM units to capture temporal dependencies in the target motion.  {The network is trained on fixed-length sequences but can be applied recursively at inference time, providing an attitude estimate at each frame while exploiting temporal information through its internal memory. The estimate associated with the current frame corresponds to the latest LiDAR acquisition.}

\begin{figure*}[h!]
    \centering
    \includegraphics[width=0.9\textwidth]{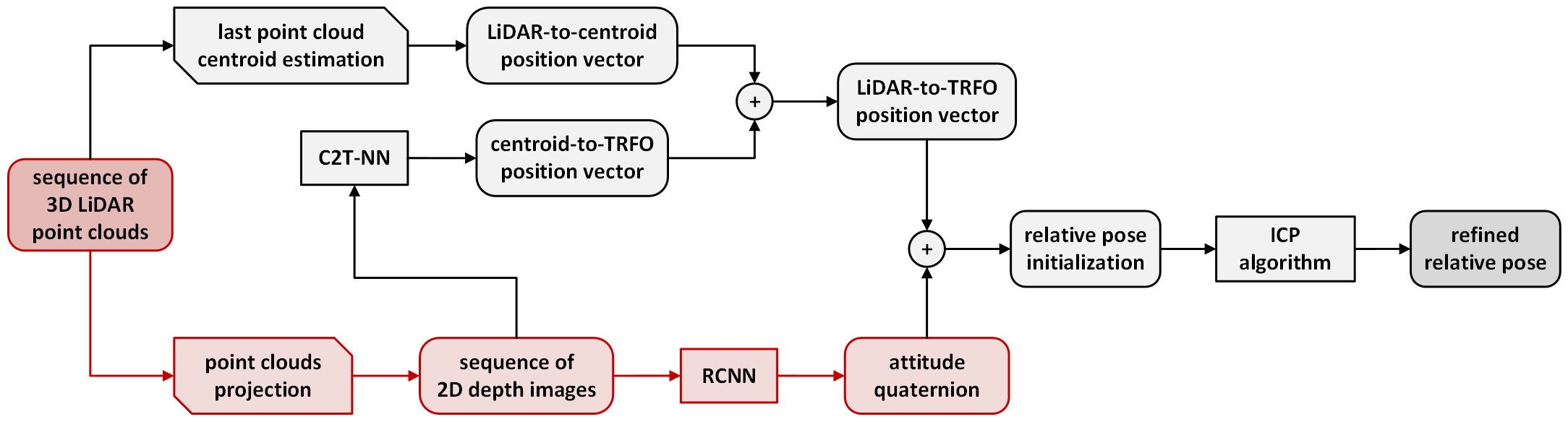}
    \caption{End-to-end pipeline for relative pose estimation from 3D point clouds. The steps of the method under study are highlighted in red.}
    \label{fig:pipeline}
\end{figure*}

Once trained for a given target, the RCNN estimates the orientation of the TRF with respect to the LRF. The complete processing pipeline for relative pose estimation is schematized in Fig.~\ref{fig:pipeline}, where the modules addressed in this paper are highlighted in red.

\subsection{Training parameters and dataset configuration tuning}
\label{sec:tuning}

The CNN and RCNN models are trained using datasets generated as described in Section 2.
 {Each neural network is trained on data generated from a single spacecraft model, and separate networks are trained for each spacecraft geometry considered in this study.}
Training is performed over multiple sessions in order to progressively refine the network parameters. The main training settings adopted for CNN pretraining and RCNN training are summarized in Table~\ref{tab:training_summary}.

\begin{table}[h]
    \centering
    
    \begin{tabular}{lcc}
    \toprule
        \textbf{Parameter} & \textbf{CNN} & \textbf{RCNN} \\
        \midrule
        MiniBatchSize & 64--512 & 32 \\
         {LearnRate} & $10^{-3}$--$10^{-5}$ & $10^{-3}$ \\
        Optimizer & Adam & Adam \\
        GradientThreshold & 1 & 1 \\
        ValidationFrequency & 10 & 10 \\
        Epochs per session & 200 & 100 \\
        Dataset size per session & 10,000 images & 3,000 sequences \\
        Frames per sequence & -- & 10 \\
        Convolution kernel size & $3\times3$ & $3\times3$ \\
    \bottomrule
    \end{tabular}
    
    \vspace{7pt}
    \caption{Summary of training parameters for CNN pretraining and RCNN training}
    \label{tab:training_summary}
\end{table}

{Following an approach similar to ChiNet~\citep{Rondao}, the training procedure is divided into two phases. The CNN is first pretrained on single-frame depth images, after which its weights are transferred to the RCNN.} 
 {During CNN pretraining, the mini-batch size and learning rate are progressively adjusted across successive training sessions within the ranges reported in Table \ref{tab:training_summary}, while remaining constant within each session. All other parameters are kept fixed.}
 {During RCNN training, the pretrained CNN layers are assigned a strongly reduced learning rate of $10^{-9}$, effectively freezing their weights, while only the newly introduced LSTM layer and the output layer are trained using a learning rate of $10^{-3}$ as indicated in Table \ref{tab:training_summary}.}
For both networks, after each training session the epoch yielding the best performance is selected, and the corresponding network learnable parameters are used to initialize the subsequent training session. 

Architectural and dataset parameters were selected based on a tuning analysis conducted on the double-symmetry spacecraft shown in Fig.~\ref{fig:NEAR}, which represents the most challenging scenario in terms of attitude ambiguity. By adopting this target for parameter selection, the resulting configuration is expected to remain robust under visually ambiguous conditions.

For CNN pretraining, a fixed dataset size is employed to ensure consistent feature learning across training sessions. For RCNN training, temporal datasets composed of fixed-length image sequences are adopted to provide sufficient temporal context for pose estimation. The final training setup is selected as a compromise between estimation accuracy and model complexity, and is applied consistently across all experiments to enable a fair performance comparison.

\section{Results}
\label{sec:results}

The results presented in this section were obtained by simulating, in a Matlab environment, the pipeline highlighted in red in Fig. \ref{fig:pipeline}.
During simulations, realistic point cloud sequences were generated and projected as described in Section \ref{Dataset}. The objective is to assess the accuracy and sensitivity of the proposed navigation strategies under varying conditions. All analyses discussed below concern the testing phase, performed on previously trained networks to evaluate their performance on the considered scenarios. 
The simulations aim to examine the influence of target angular velocity, relative distance, spacecraft symmetry, and image quality degradations on the neural network’s ability to produce reliable outputs as described below. 

\begin{figure}[h]
    \centering
    \begin{subfigure}[b]{0.32\columnwidth}
        \centering
        \includegraphics[width=\textwidth]{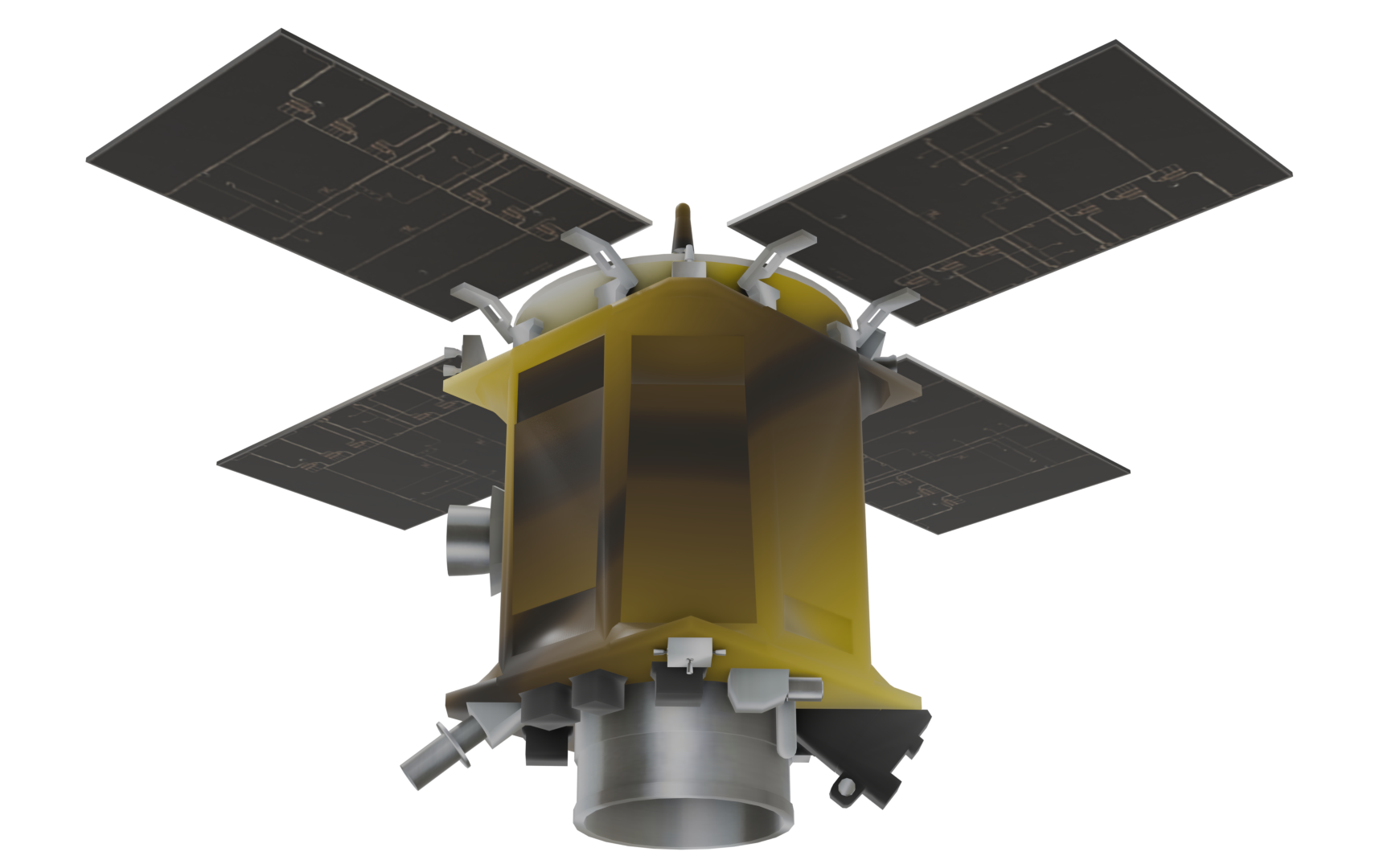}
        \caption{NEAR Shoemaker}
        \label{fig:NEAR}
    \end{subfigure}
    \hfill
    \begin{subfigure}[b]{0.32\columnwidth}
        \centering
        \includegraphics[width=\textwidth]{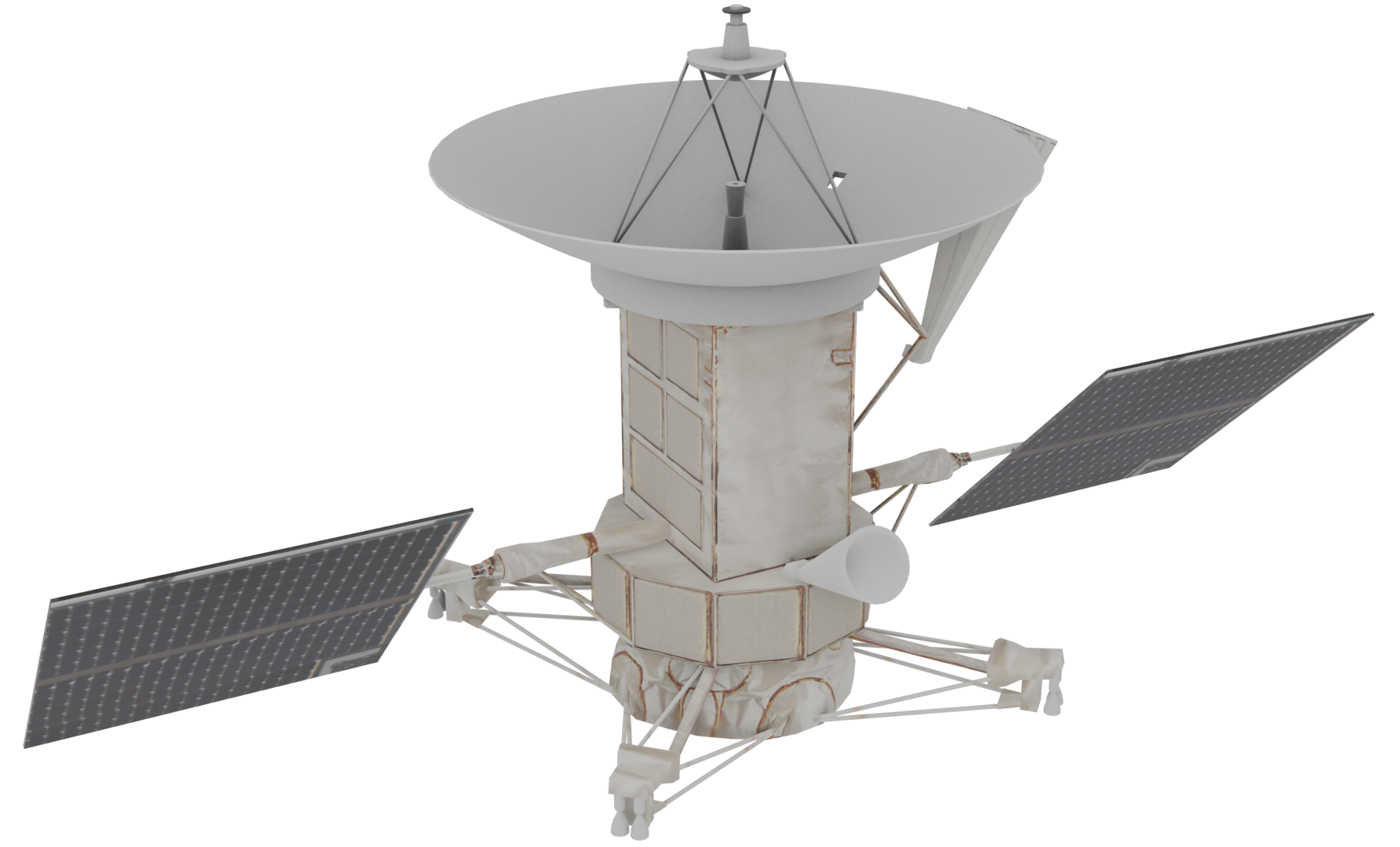}
        \caption{Magellan probe}
        \label{fig:Magellan}
    \end{subfigure}
    \hfill
    \begin{subfigure}[b]{0.32\columnwidth}
        \centering
        \includegraphics[width=\textwidth]{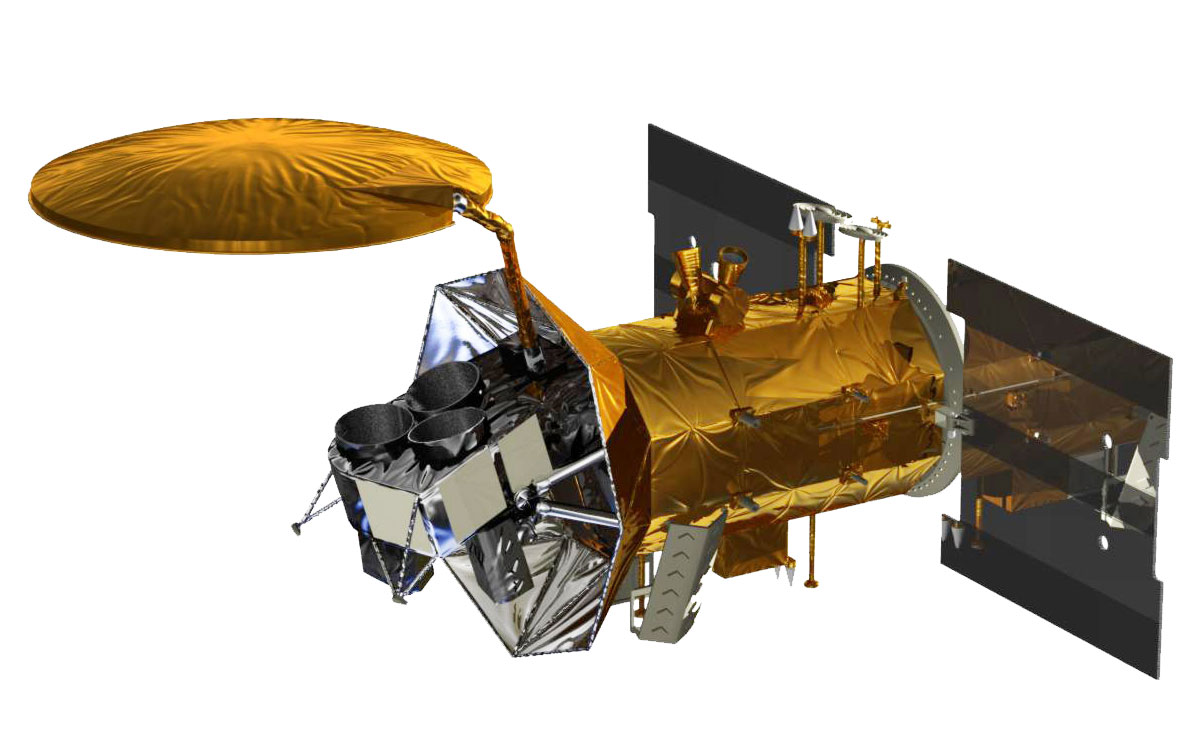}
        \caption{SAC-D/Aquarius}
        \label{fig:Aquarius}
    \end{subfigure}
    \caption{Conceptual rendering of the considered spacecraft. Spacecraft (\subref{fig:NEAR}) exhibits a double quasi-symmetry, spacecraft (\subref{fig:Magellan}) presents a single approximate symmetry, and spacecraft (\subref{fig:Aquarius}) has no symmetries. Credit: NASA}
    \label{fig:spacecrafts}
\end{figure}

\begin{figure}[h]
    \centering
    \begin{subfigure}[b]{0.49\columnwidth}
        \centering
        \includegraphics[width=\textwidth]{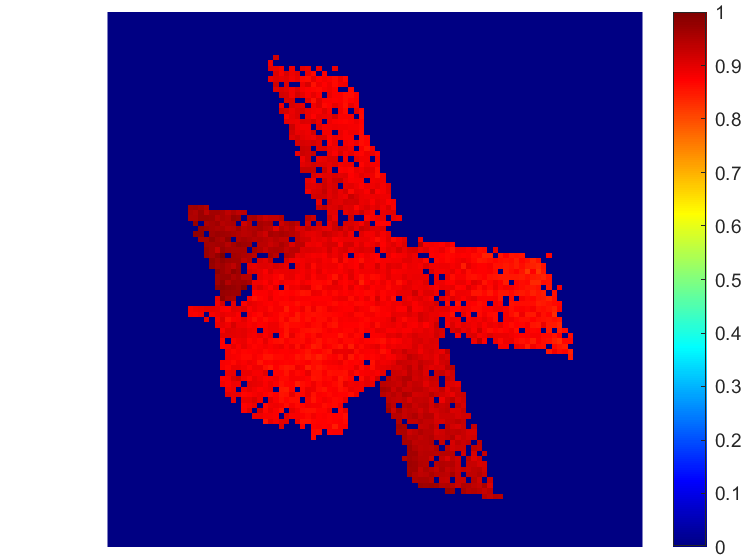}
        \caption{Resolution = FOV/128}
    \end{subfigure}
    \hfill
    \begin{subfigure}[b]{0.49\columnwidth}
        \centering
        \includegraphics[width=\textwidth]{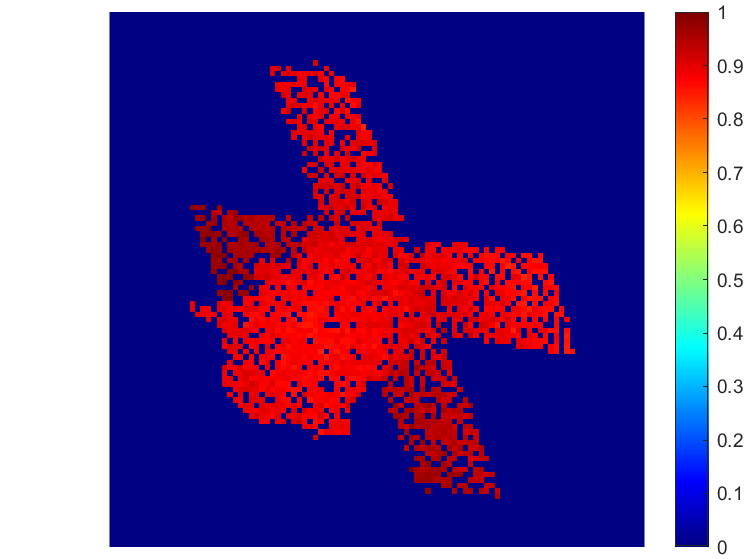}
        \caption{Resolution = FOV/100}
    \end{subfigure}
    \caption{Depth images showing NEAR Shoemaker, generated with two different angular resolutions}
    \label{fig:angular_resolution}
\end{figure}

\begin{figure}[h]
    \centering
    \begin{subfigure}[b]{0.49\columnwidth}
        \centering
        \includegraphics[width=\textwidth]{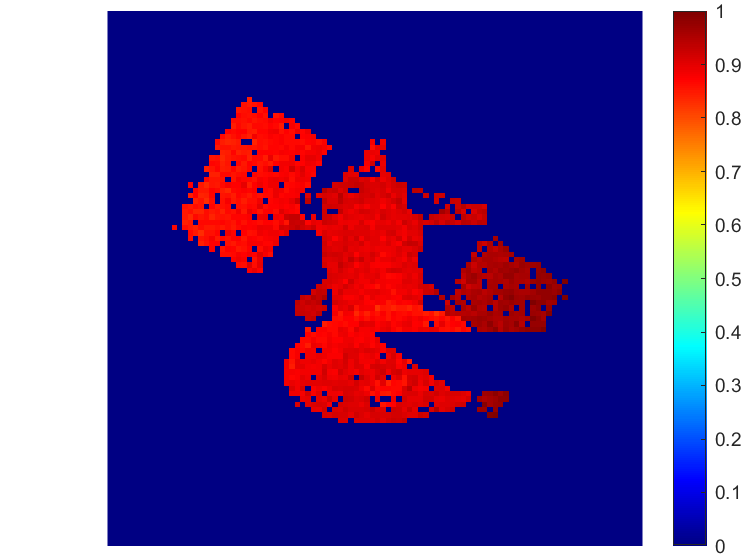}
        \caption{With partial occlusion}
        \label{fig:Magellan_occlusion}
    \end{subfigure}
    \hfill
    \begin{subfigure}[b]{0.49\columnwidth}
        \centering
        \includegraphics[width=\textwidth]{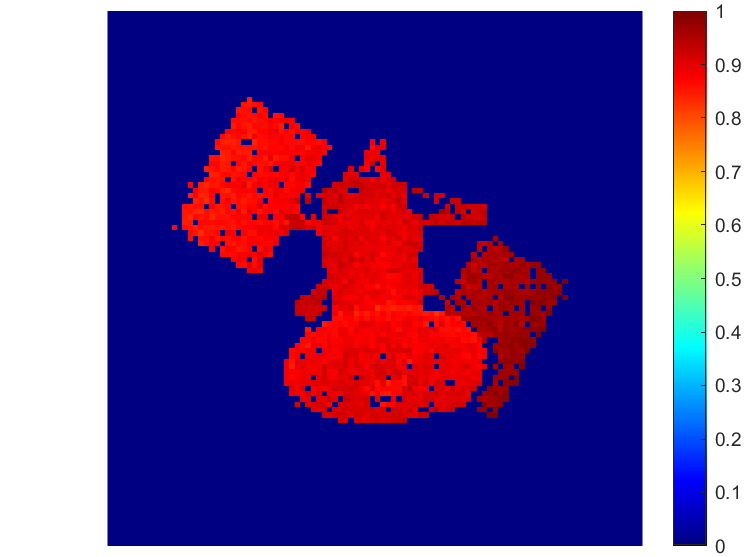}
        \caption{Without occlusion}
        \label{fig:Magellan_no_occlusion}
    \end{subfigure}
    \caption{Example of an occlusion in a frame showing the Magellan space probe}
    \label{fig:occlusions}
\end{figure}

Angular velocity profiles were generated with random components along all three body axes, with a total angular rate magnitude varying from 0$^\circ$/s to 20$^\circ$/s in steps of 2.5$^\circ$/s. Furthermore, the target distance was varied between 20~m and 100~m in increments of 10~m.
Then, three spacecraft geometries were considered to span a range of symmetry properties. The models, illustrated in Fig. \ref{fig:spacecrafts}, correspond to a two-axis symmetric body (NEAR Shoemaker, Fig. \ref{fig:spacecrafts}\subref{fig:NEAR}), a single-axis symmetric probe (Magellan, Fig. \ref{fig:spacecrafts}\subref{fig:Magellan}), and an asymmetric spacecraft with no symmetry axes (SAC-D/Aquarius, Fig. \ref{fig:spacecrafts}\subref{fig:Aquarius}). This selection enabled a systematic assessment of the networks’  {capability to handle} orientation-induced appearance ambiguities.
Finally,  {the sensitivity} to input image quality is evaluated by subjecting the simulated sensor images to three degradation conditions: (i) nominal, high-quality imagery representative of ideal sensor performance with a resolution of FOV/128 (Fig. \ref{fig:angular_resolution}a); (ii) moderate degradation, including a reduced angular resolution (FOV/100) and the random removal of 10\% of the input points to emulate missed LiDAR returns caused by surface material properties, low reflectivity, or unfavorable incidence angles (Fig. \ref{fig:angular_resolution}b);
and (iii) severe degradation, where portions of the input image were intentionally obscured to simulate realistic visual occlusions arising from spacecraft appendages (e.g., robotic arms, antennas, or solar panels) intermittently entering the LiDAR field of view during on-orbit proximity operations. In this case, the resolution is considered to be equal to FOV/128. 
An example of the resulting occlusion effect is shown in Fig.~\ref{fig:occlusions}. All test conditions employ the same LiDAR simulation framework described in Section~\ref{Dataset} and the same target CAD models.
 {The adopted simulator also introduces sporadic range errors exceeding twice the nominal sensor accuracy, thus implicitly accounting for isolated outliers.}
More specifically, the networks were trained under nominal resolution conditions (FOV/128) and degraded conditions (FOV/100 and missed returns), whereas occlusions and tumbling motion are introduced exclusively during testing.

To summarize the testing methodology, the RCNN described in Section~\ref{sec:RCNN} was evaluated through 900 Monte Carlo simulations for each combination of spacecraft geometry (Fig. \ref{fig:spacecrafts}) and image-quality condition (Figs. \ref{fig:angular_resolution} and \ref{fig:occlusions}). Within each set, target distance and angular velocity were randomized within the previously defined ranges to capture performance variability and failure modes. 
Furthermore, the RCNN was compared to the CNN described in Section~\ref{sec:CNN} to assess the performance improvement achieved through the recurrent design. The evaluation was conducted under the nominal high-quality condition (FOV/128).
Finally, an angular error threshold of 20° was defined in the numerical simulations to determine convergence success for the proposed attitude initialization methods.  
This conservative choice was made to ensure reliable convergence of the subsequent ICP-based pose refinement, which can typically converge even from initial errors up to 30°. In the following, the final angular error achieved by the proposed methods is referred to as the \textit{initialization error}, as it represents the initial misalignment provided to the subsequent ICP stage.

\begin{table}[h]
    \centering
    \resizebox{\columnwidth}{!}{%
    \begin{tabular}{ccccc}
    \toprule
         \textbf{Spacecraft} & \multicolumn{2}{c}{\textbf{RCNN}} & \multicolumn{2}{c}{\textbf{CNN}} \\
         \cmidrule(lr){2-3} \cmidrule(lr){4-5}
         & mean ang. err. & $\leq$20° & mean ang. err. & $\leq$20° \\
         \midrule
         NEAR Shoemaker & 12.69° & 88\% & 40.94° & 45\%\\
         Magellan probe & 3.99° & 99\% & 6.36° & 94\%\\
         SAC-D/Aquarius & 3.09° & 100\% & 5.31° & 95\%\\
    \bottomrule
    \end{tabular}
    } 
    \vspace{5pt}
    \caption{ {Performance} of RCNN in Fig. \ref{fig:RCNN_architecture} and CNN in Fig. \ref{fig:CNN_architecture} for sensor resolution of FOV/128
    \label{tab:RCNN_performance}}
\end{table}

\begin{table}[h]
    \centering
    \begin{tabular}{ccc}
    \toprule
         \textbf{Spacecraft} & \textbf{mean ang. err.} & \textbf{test with err. $\leq$ 20°}\\
         \midrule
         NEAR Shoemaker & 14.09° & 84\%\\
         Magellan probe & 4.57° & 100\%\\
         SAC-D/Aquarius & 3.86° & 100\%\\
    \bottomrule
    \end{tabular}
    \vspace{7pt}
    \caption{Performance of the RCNNs on test datasets with degraded resolution}
    \label{tab:RCNN_performance_degraded_resolution}
\end{table}

\begin{table}[h]
    \centering
    \begin{tabular}{ccc}
    \toprule
         \textbf{Spacecraft} & \textbf{mean ang. err.} & \textbf{test with err. $\leq$ 20°}\\
         \midrule
         NEAR Shoemaker & 22.86° & 64\%\\
         Magellan probe & 13.89° & 83\%\\
         SAC-D/Aquarius & 12.82° & 86\%\\
    \bottomrule
    \end{tabular}
    \vspace{7pt}
    \caption{Performance of the RCNNs in the presence of occlusions}
    \label{tab:RCNN_performance_occlusions}
\end{table}

\begin{table}[h]
    \centering
 
    \begin{tabular}{ccc}
    \toprule
         \textbf{Spacecraft} & \textbf{mean ang. err.} & \textbf{test with err. $\leq$ 20°}\\
         \midrule
         NEAR Shoemaker & 13.16° & 85\% \\
         Magellan probe & 4.86° & 99\% \\
         SAC-D/Aquarius & 3.97° & 99\% \\
    \bottomrule
    \end{tabular}
    
    \vspace{7pt}
    \caption{Performance of the RCNNs in tumbling conditions with sensor resolution of FOV/128}
    \label{tab:RCNN_performance_tumbling}
\end{table}

 {The attitude initialization accuracies achieved by} the RCNN (Fig. \ref{fig:RCNN_architecture}) and CNN (Fig. \ref{fig:CNN_architecture}) architectures under the nominal high-quality condition (FOV/128) are reported in Table~\ref{tab:RCNN_performance}. 
Both models are evaluated under identical dynamic and sensing conditions, including the same spacecraft geometries, distance ranges, angular velocity distributions, and LiDAR simulation parameters. 
The table lists, for each spacecraft geometry, the mean angular error, corresponding to the average initialization error, and the percentage of simulations with a final angular error below the 20$^\circ$ convergence threshold. 
 {The results show that the RCNN achieves lower attitude errors and higher convergence rates than the CNN under the considered test conditions.} For the NEAR Shoemaker spacecraft, which exhibits two symmetry axes, the percentage of successful convergences nearly doubles, increasing from 45\% to 88\%. Significant improvements are also observed for the Magellan and SAC-D/Aquarius spacecraft, where the mean angular error is almost halved.
 {This improvement is related to the exploitation of temporal information through recurrent processing, although the different data structures used for training (single-frame images for CNN and temporally correlated sequences for RCNN) may also contribute to the observed performance gap. Therefore, the results should be interpreted as the benefit of the complete sequence-based recurrent approach rather than as an isolated effect of the recurrent architecture alone.}

Then, the influence of image quality on the RCNN performance can be assessed by comparing Tables~\ref{tab:RCNN_performance}–\ref{tab:RCNN_performance_occlusions}. Results show that a moderate degradation of the input resolution (Table~\ref{tab:RCNN_performance_degraded_resolution}, FOV/100) does not significantly affect the estimation accuracy or convergence percentage, indicating that the network maintains stable performance even under reduced visual detail.
Conversely, the presence of occlusions leads to a noticeable performance drop, with a reduction of successful convergences of about 20\% and an average angular error increase of approximately 10$^\circ$ across all spacecraft geometries. This degradation can be attributed to the partial loss of key geometric features in the input projections, which limits the network’s ability to estimate the target attitude.

The RCNNs, which are trained under constant angular velocity conditions, are subsequently evaluated on  
tumbling target scenarios. The evaluation is conducted on 1,000 test sequences. The inertia tensors are not available online, as a consequence they are obtained through the CAD files considering a mean density calculated from the nominal mass and volume of each spacecraft.

\begin{flalign}
\mathbf{I}_{NEAR\; Shoemaker} & =
\begin{bmatrix}
372.38 & -2.86 & -2.95\\
-2.86 & 372.45 & -4.42\\
-2.95 & -4.42 & 316.17
\end{bmatrix} \text{kg}\cdot m^2
&
\end{flalign}

\begin{flalign}
\mathbf{I}_{Magellan} & =
\begin{bmatrix}
3266.61 & 2.62 & 36.17\\
2.62 & 4772.03 & -42.65\\
36.17 & -42.65 & 2897.59
\end{bmatrix} \text{kg}\cdot m^2
&
\end{flalign}

\begin{flalign}
\mathbf{I}_{\text{Aquarius}}&=
\begin{bmatrix}
2644.24 & 972.97 & -2.10\\
972.97 & 1329.64 & 11.49\\
-2.10 & 11.49 & 3086.41
\end{bmatrix} \text{kg}\cdot m^2
&
\end{flalign}

 {The results reported in Table~\ref{tab:RCNN_performance_tumbling} show only limited performance degradation under the considered tumbling conditions.} 
Compared to Table \ref{tab:RCNN_performance}, the mean angular error increases by less than $1^\circ$ for all targets, while the success rate remains nearly unchanged. The largest reduction, approximately 3\%, is observed for the NEAR Shoemaker, which is particularly challenging because of its high degree of symmetry.
 {To further investigate the sensitivity of the RCNN to rotational dynamics not represented during training,} the NEAR Shoemaker inertia tensor was artificially modified by halving $I_{zz}$. This produces stronger inertia coupling effects  {and a larger deviation from the constant-angular-velocity assumption.} In this non-physical stress test, the mean angular error increases to $17.21^\circ$, while the success rate decreases to 74\%.
 {Overall, the proposed RCNN achieves satisfactory results for both spinning targets and tumbling scenarios with nominal inertia properties. However, the modified-inertia test highlights that generalization to arbitrary tumbling dynamics cannot be fully guaranteed. 
Pronounced tumbling motions generate temporal patterns that deviate from the constant-angular-velocity sequences used for training, thereby reducing the effectiveness of the recurrent module.}

\begin{figure}[h]
    \centering
    \includegraphics[width=0.95\linewidth]{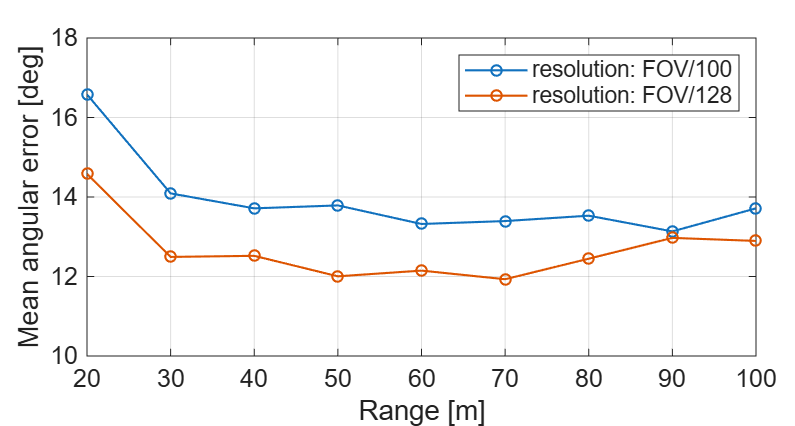}
    \caption{Performance variation with range for the RCNN trained on the NEAR Shoemaker space probe}
    \label{fig:performance_range}
\end{figure}

\begin{figure}[h]
    \centering
    \includegraphics[width=0.95\linewidth]{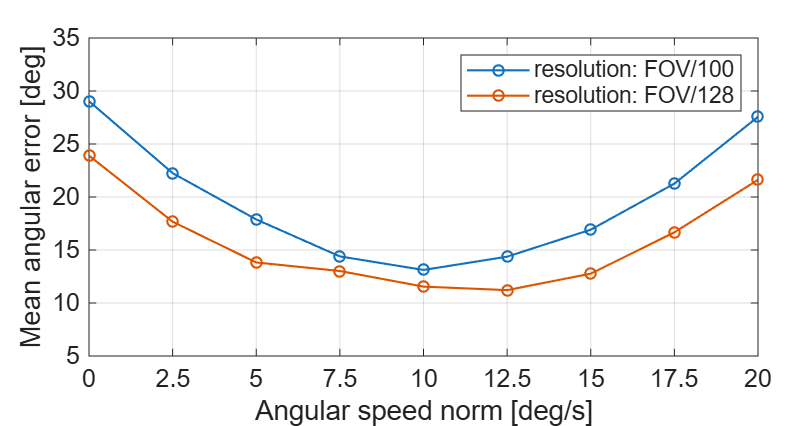}
    \caption{Performance variation with angular speed for the RCNN trained on the NEAR Shoemaker space probe}
    \label{fig:performance_angular_speed}
\end{figure}

The performance of the RCNN as a function of target distance and angular velocity is quantified in Figs.~\ref{fig:performance_range} and~\ref{fig:performance_angular_speed}, respectively. For this analysis, the NEAR Shoemaker spacecraft was selected, as it represents the most challenging case due to its two symmetry axes, and tests were performed under both FOV/128 and FOV/100 conditions.
Fig.~\ref{fig:performance_range} shows that the RCNN maintains consistent performance across the considered range of distances, highlighting the effectiveness of the voxelization strategy and the representativeness of the training datasets. Specifically, the results were obtained from datasets generated at fixed ranges, allowing an assessment of potential correlations between target distance and performance degradation.

Then, Fig.~\ref{fig:performance_angular_speed} illustrates the variation of the initialization error with respect to the target's angular velocity norm, and was obtained using a training dataset in which the angular velocity vectors were generated by sampling their components from uniform distributions. This procedure results in a non-uniform distribution of angular velocity magnitudes, with a higher concentration of samples around intermediate speeds and fewer samples at the boundaries of the considered range.

\begin{figure}[h]
    \centering
    \includegraphics[width=0.95\linewidth]{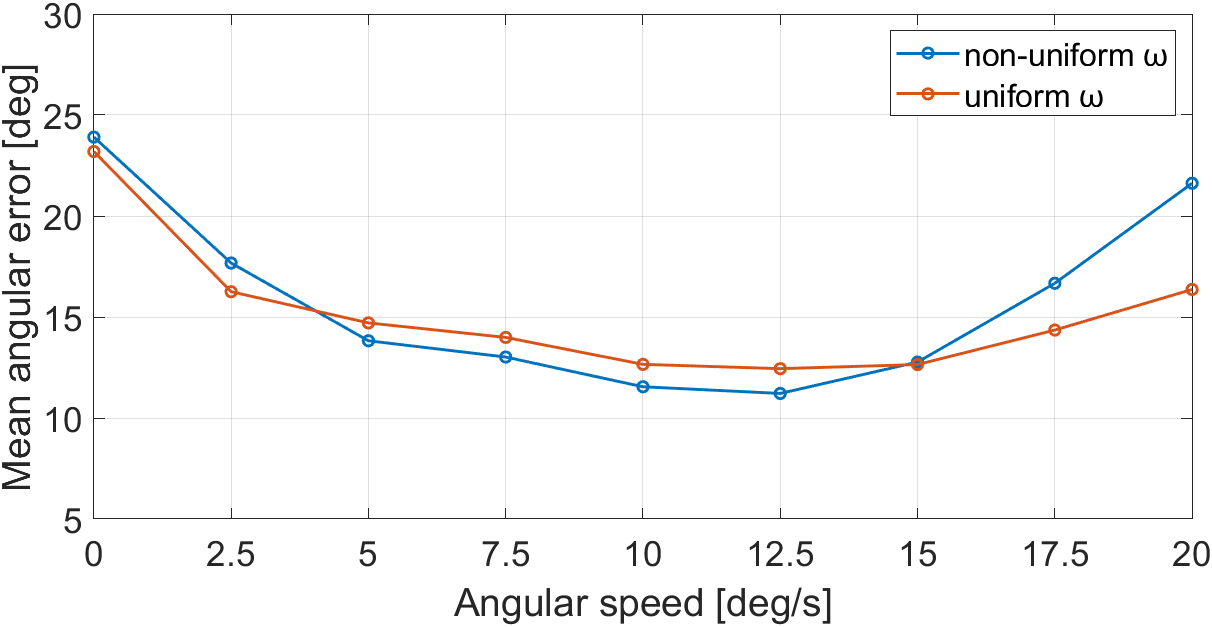}
    \caption{ {Performance vs. angular speed for RCNNs trained on uniform/non-uniform angular speed distributions}}
    \label{fig:performance_angular_speed_uniform_omega}
\end{figure}

In order to better interpret the dependence of the trend reported in Fig.~\ref{fig:performance_angular_speed} on the composition of the training dataset, an additional training strategy based on a uniform distribution of angular velocity magnitudes between 0°/s and 20°/s was considered. In this case, the RCNN was trained on a sequential dataset of the NEAR Shoemaker, selected due to its high symmetry, generated using uniformly distributed angular velocity norms.
Fig.~\ref{fig:performance_angular_speed_uniform_omega} compares the estimation accuracy obtained using networks trained with non-uniform (blue line) and uniform (orange line) angular velocity distributions, both evaluated under nominal sensor conditions.  
 {The comparison shows that, for the original non-uniform training strategy, the estimation error exhibits higher values at the boundaries of the considered angular velocity range, particularly for angular velocities below  $5^\circ/s$ and above  $15^\circ/s$, which are less represented in the training dataset. Conversely, intermediate angular velocities benefit from a denser representation in the training data.
However, both training approaches exhibit qualitatively similar trends, characterized by improved accuracy at intermediate angular velocities and increased errors towards the boundaries of the considered range.
These results indicate that the sampling strategy affects the performance, particularly in regimes that are less represented in the training dataset, while the persistence of similar trends across the two strategies suggests that the estimation accuracy is also influenced by the characteristics of the attitude estimation problem at different angular velocities.}

\subsection{Comparison with State-of-the-Art Approaches}

This subsection compares the proposed RCNN with existing LiDAR-based deep learning approaches for spacecraft pose initialization. The proposed comparison is based on key criteria that characterize the problem, such as operational range, target symmetries, presence of occlusions, and management of temporal information.

\subsubsection{Operational Range}

Most existing deep learning-based LiDAR pose estimation approaches, such as \citep{Leo1, zhang, Piccinin, Hashimoto}, report evaluation ranges below 20\,m, with most experiments conducted at distances smaller than 5\,m. In contrast, \citep{Enrique} evaluates attitude initialization at a fixed relative distance of 100\,m.

The proposed RCNN is evaluated over a wide range of relative chaser--target distances, from 20\,m to 100\,m, independently of target shape or rotational motion. The results show that attitude estimation accuracy remains nearly constant across the entire range,  {showing limited sensitivity} to distance-related point cloud degradation and extending the applicability domain of the proposed approach.

\subsubsection{Target Symmetries}

Several existing works, including \citep{zhang, Enrique}, consider only asymmetric target geometries. When evaluated on the asymmetric SAC-D/Aquarius satellite, the proposed RCNN achieves an improvement of approximately 28\% in attitude estimation accuracy compared to the best results reported in these works.

Targets exhibiting multiple symmetries are instead addressed in \citep{Piccinin, Leo1}. In these scenarios, the proposed method shows a higher mean attitude estimation error (approximately 10$^\circ$), while maintaining a high convergence rate exceeding 88\%. This degradation is primarily attributable to the 20 times larger operational range of the proposed approach, which results in noisier and more ambiguous point cloud data. Unlike \citep{Leo1}, the proposed method does not rely on satellite-specific geometric assumptions, enabling applicability to arbitrary target geometries.

\subsubsection{Occlusions}
The impact of partial occlusions within the LiDAR field of view has received limited attention in the existing literature, which typically evaluates pose estimation performance under ideal visibility conditions. In this work, the proposed RCNN is evaluated under partial occlusion scenarios, showing limited degradation in attitude estimation accuracy and a consistently high convergence rate.

\subsubsection{Temporal Information}
Unlike prior works, the proposed RCNN explicitly accounts for the temporal evolution of the target's attitude through the integration of LSTM units, enabling the exploitation of temporal correlations across consecutive LiDAR frames and improving estimation consistency in dynamic scenarios.
While recurrent architectures have been widely adopted in RGB-based pose estimation approaches \citep{Rondao, d'amico2, d'amico1}, their extension to LiDAR-based spacecraft pose estimation remains largely unexplored. This work extends the RCNN paradigm to LiDAR data, leveraging its inherent advantages in terms of operational range, reduced sensitivity to lighting conditions, and direct depth information.

 {\subsection{Limitations and Practical Considerations}}

 {Although the proposed framework achieves strong performance under the considered simulation conditions, some factors should be taken into account when assessing its applicability to real-world scenarios.}

First, real-world datasets may introduce additional challenges due to the so-called \textit{domain gap}, as discussed in \citep{Leo1}.  {Furthermore, another important practical consideration is the inference runtime of the proposed networks. Since no flight-qualified onboard computer was available, all runtime measurements were performed on an 11th Gen Intel® Core™ i5-1135G7 CPU operating at 2.40 GHz. Under these conditions, the RCNN required approximately 24 ms per frame, while the CNN processed a single depth image in about 4.8 ms. These measurements should not be interpreted as evidence of real-time capability on space-grade hardware. Rather, they provide an indication of the computational cost of the proposed approach and allow a qualitative comparison with previous studies. For example, \citep{Leo1} reported an inference time of approximately 33 ms for a CNN executed on an Intel® Xeon W-2135 CPU. Direct comparisons, however, remain approximate because of differences in hardware platforms, software implementations, network architectures, and input representations. From the perspective of the simulated sensing pipeline, the measured inference times are substantially shorter than the LiDAR sampling period of 2.7 s. Nevertheless, establishing the suitability of the proposed framework for onboard real-time operation would require dedicated benchmarking on representative flight processors or other space-qualified embedded computing platforms.
}

\section{Conclusion}

This work presented an RCNN-based method for coarse attitude estimation of known spacecraft from LiDAR data. 
 {By exploiting temporal information from consecutive observations, the proposed approach achieved lower estimation errors than the considered CNN baseline across the tested scenarios.  Since the CNN and RCNN were trained using datasets with different temporal structures, the observed performance improvements should be interpreted as the effectiveness of the complete sequence-based approach.} The proposed framework maintained consistent performance under variations in spacecraft geometry, relative distance, and image quality,  {while showing accuracy reduction for low and high angular velocities values.} 
 {Furthermore, the generalization to arbitrary tumbling motions may be limited by the adopted training conditions, as highlighted by the tumbling stress-test analysis.} 
 {From a computational perspective, the measured inference times are lower than the LiDAR acquisition period considered in this study on the tested commercial platform. However, this result does not imply real-time suitability for flight implementation, which requires dedicated evaluation on representative onboard hardware.}
Future work will focus on validating the approach on real datasets, investigating robustness to larger disturbances and outliers, and extending the framework to additional relative navigation scenarios, including asteroid-relative operations.

 {\section{Acknowledgments}
This publication is part of the project PNRR-NGEU which has received funding from the MUR – DM 117/2023 and DM 630/2024.
\begin{figure}[h!]
    \centering
    \includegraphics[width=1\linewidth]{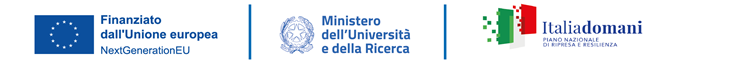}
\end{figure}}


 \bibliographystyle{elsarticle-num} 
\bibliography{ifacconf}             
\end{document}